\documentclass[preprint]{aastex}
\usepackage{graphicx}

\newcommand{\be}{\begin{equation}}
\newcommand{\ee}{\end{equation}}

\newcommand{\ba}{\begin{eqnarray}}
\newcommand{\ea}{\end{eqnarray}}

\newcommand\lo{\mathrel{\raise.3ex\hbox{$<$}\mkern-14mu\lower0.6ex\hbox{$\sim$}}}
\newcommand\go{\mathrel{\raise.3ex\hbox{$>$}\mkern-14mu\lower0.6ex\hbox{$\sim$}}}

\begin{document}
\title{Comment on ''Gamma-ray burst early afterglows: 
reverse shock emission from an arbitrarily magnetized ejecta'' 
by Zhang \& Kobayashi (2004)}
\author{
Maxim   Lyutikov}
\affil{University of British Columbia, 6224 Agricultural Road,
Vancouver, BC, V6T 1Z1, Canada}

\begin{abstract}
Zhang \& Kobayashi (2004) attempted to calculate early afterglow
 emission from a system of forward and reverse shocks 
in GRB outflows for the case of   magnetized ejecta.
We point out a  fundamental error in the underlying 
 dynamical model.
According to the  authors,   energy and momentum carried 
by   the magnetic field of the ejecta are not
transfered to the forward shock. This is an incorrect assumption that invalidates
the results.
\end{abstract}

Observations of early afterglows, almost coincident with the prompt phase, may serve
as  a simple test of ejecta content (Lyutikov 2004). This is expected to be probed
by  incoming data from Swift satellite. One of the principal issues  at stake is 
what fraction of the  energy released by the central  source
is carried by magnetic field.
Zhang \& Kobayashi (2004) 
 attempted to model emission from a system of reverse and forward shocks
for highly magnetized ejecta. Here we point out a fundamental error in their calculations. 
When considering interaction of magnetized ejecta with external medium, Zhang \& Kobayashi (2004)
assumed ``inability of tapping the Poynting flux energy [by] the forward shock``. 
Thus, for highly magnetized ejecta only kinetic part of energy was assumed to be given 
to the forward
shock. The fate of magnetic energy was not discussed; it is implicitly assumed that
magnetic field energy disconnects from the flow when it reaches a contact discontinuity separating 
ejecta and circomstellar medium, thus behaving as radiation.
In their own words, ``only the kinetic energy of the baryonic component
[] defines the energy that interacts with the ambient
medium''. This point of view is incorrect since the energy
of the  Poynting flux is also transfered to the ejecta.
In fact, magnetized ejecta may transfer energy and momentum 
to the circomstellar medium even more efficiently that unmagnetized one.
For unmagnetized ejecta, the  two relativistic fluids couple through
development of  two steam instability. In case of highly magnetized ejecta,
particles from circomstellar medium enter magnetic field of the ejecta,
complete half a turn and are reflected back, gaining 
in the laboratory frame energy $\propto \Gamma_0^2$, where $\Gamma_0$ is a Lorentz factor
of the contact discontinuity. As a result,
magnetic field of the ejecta makes a $pdV$ work on circomstellar medium.
This type of interaction is well understood in space physics:
in case of solar wind--Earth magnetosphere interaction a  two gyro-radius displacement
of reflected particles  along the contact discontinuity 
creates a so called Chapman-Ferraro current which shields  the Earth magnetic dipole.

The underlying dynamical model of Zhang \& Kobayashi (2004)
underestimates the energy of the forward shock by 
a factor of $\sigma$, which can be as large as $10^3$ in their calculations. 
The error of Zhang \& Kobayashi (2004)
stems from the incorrect assumption that ejecta will start to decelerate
when the swept mass becomes of the order of the ejecta mass, thus neglecting inertia
associated with magnetic field. Under ideal MHD, when stress-energy tensor is diagonalizable,
magnetic field has effective rest-frame inertial density $\rho_{MHD} = b^2 /(8 \pi c^2)$ where $b$ is magnetic field
in the rest-frame (frame  where  electric field is zero). 
 Highly magnetized ejecta, parametrized by the ratio of Poynting to particle 
fluxes $\sigma$ and  
moving with Lorentz factor $\Gamma_0$,  starts to  decelerate
when a large fraction of ejecta energy has been transfered to the forward shock:
\be
r_{dec} \sim \left( { E_0 \over \Delta \Omega \rho c^2 \Gamma_0^2 } \right)^{1/3}
\ee 
and not at a radius when the swept-up mass equals the ejecta particle mass
\be
r_{swept} \sim \left( { M_0 \over \Delta \Omega \rho c^2 \Gamma_0 } \right)^{1/3}
= \left( { E_K  \over \Delta \Omega \rho c^2 \Gamma_0^2 } \right)^{1/3}
\ee
Here $E_0 $ is the total energy of ejecta, $\Delta \Omega$ is solid angle
of explosion and
 $E_K=E_0/(1+\sigma)$ is the energy associated with
bulk motion of matter.
Only in the case of zero magnetization the two definitions are nearly equivalent, since
in that case $  E_0=E_K  = M_0 \Gamma_0$. For arbitrary $\sigma$
\be
{r_{swept} \over r_{dec}} = (1+\sigma)^{-1/3} \ll 1
\ee
where the last inequality applies for $\sigma \gg 1$.
Zhang \& Kobayashi (2004) ''define the deceleration
radius using $E_K$ alone [] where the
fireball collects $1/\Gamma_0$ of fireball rest mass'' (so that deceleration
radius radius is $r_{swept}$).  This is incorrect; the flow starts to decelerate 
at $r_{dec} \gg r_{swept}$.

Qualitatively, at the Blandford-McKee stage, $r> r_{dec}$, 
energy in the forward shock is determined by the total energy of the source and is virtually independent
of the  content of the ejecta, while $\Gamma \propto t^{-3/2}$ (in a constant density environment).
Thus, at these times the forward shock emission should follow
the $\sigma=0$ line.
As a result, {\it late afterglow observations cannot be used to infer ejecta content}
(see Lyutikov, 2004, for more discussion).
At smaller radii, $r < r_{dec}$, for MHD-type 
expansion, $\Gamma_0 > \sqrt{\sigma}$,
Lorentz factor is constant, while for force-free-type expansion,  $\Gamma_0 < \sqrt{\sigma}$,
Lorentz factor is decreasing but 
at a slower rate ($\propto t^{-1/2}$ in a constant density medium, 
Lyutikov, 2004).

Incorrect assumptions about dynamics of the forward shock  invalidate calculations
of emission both from reverse and forward shocks, since the two are related by jump
and continuity conditions. 


{}

\end{document}